\documentclass[a4paper,11pt]{article}
\pdfoutput=1
\usepackage{jcappub}
\usepackage{graphicx}
\usepackage{subfig}
\usepackage{amssymb}
\usepackage{amsmath}
\usepackage{amsthm}
\usepackage{color}

\title{\boldmath B-mode auto-bispectrum due to matter bounce}

\author[a,b]{Rahul Kothari,}
\author[a]{Debottam Nandi}
\affiliation[a]{Department of Physics, Indian Institute of Technology Madras, Chennai
600036, India}
\affiliation[b]{Current Affiliation: Department of Physics \& Astronomy, University of the Western Cape, Cape Town 7535, South Africa}
\emailAdd{quantummechanicskothari@gmail.com}
\emailAdd{debottam@physics.iitm.ac.in}

\abstract{Primordial Gravitational waves leave polarization imprints on the Cosmic Microwave Background (CMB). In this article, we investigate polarization bispectrum, which is also referred to as the B-mode auto bispectrum, due to a matter bounce Universe. For simplicity,
we consider a minimally coupled Einstein frame and obtain
an analytical integral expression for the bispectrum and numerically
perform the integration.  We find that the signal-to-noise
ratio is small, when compared with the same in the inflationary paradigm and hence quite difficult to detect in the future experiments. Thus a detection of tensor mode bispectrum in future will be helpful in ruling out matter bounce model.  Also,
to ease the numerical evaluation of the bispectrum, we develop and
use various techniques. We believe that these techniques can be used
in various  other  contexts.
}

\begin{document}
\maketitle
\flushbottom

\section{Introduction}
The Cosmic Microwave Background Radiation (CMB) is considered to be
a relic of events happened in the early Universe. Primordial fluctuations
generated in the early Universe leave imprints on the CMB. Thus CMB
helps us decipher the nature of these primordial perturbations.

These perturbations can be related to the spherical harmonic coefficients
of  CMB field through transfer functions. Thus we can evolve the
primordial perturbations, starting from the beginning to the recombination
and then to the present epoch. With the help of the correlation function
of these harmonic coefficients, we can infer about the state of the
early universe. 

CMB field, in addition to being statistically isotropic,
is also assumed to be Gaussian to a large extent. Thus, all the statistical
information is contained in the two point correlations of its harmonic
coefficients. This is because any higher even order correlations can
be expressed in terms of the  two point correlations using Wick's Theorem.
Moreover, all odd correlations turn out to be zero. However, in the
presence of non-gaussianities in the primordial fluctuations, odd
correlations are nonzero as well. The study of these primordial non-gaussianities
provides more information about the physics of the early universe and therefore
it helps to put tighter constraints on early Universe models.

Within the framework of the standard model of cosmology,
inflationary paradigm is the most remarkable one \cite{Mukhanov1990,Bassett2005,Sriramkumar2009,Baumann2009,Linde2014}.
Its success is not only based on providing a nearly scale-invariant power
spectra as required by observations, but also in solving the horizon
and flatness problems. However, even with tighter constraints obtained with the help of the non-gaussian primordial spectra, we are unable
to rule out a significant number of models within the inflationary
paradigm \citep{Martin2010,Martin2013a,Martin2013b,Martin2014,Gubitosi2015}.
Therefore, there is a growing interest in finding new alternatives
to inflation. One such popular paradigm is bounce \citep{Novello2008,Cai2014,Battefeld2014,Lilley2015,Ijjas2015,Brandenberger2016}.

Bouncing cosmologies have been present in the scientific literature
since the late 70's \citep{Novello2008}. Bouncing models represent a
situation where the Universe initially undergoes a period of contraction
until the scale factor reaches a minimum, after which it transits to
the expanding phase. Similar to inflation, these models can also solve the horizon and the flatness problems, and at the same time, can provide (near) scale-invariant spectra as required by observations. One of the most popular models of such bouncing scenarios is the {\textit{matter bounce}} Universe \citep{Starobinsky1979,Finelli2001,Raveendran2017}.

In order to find the correct model of the early Universe, we need to compare both bounce and inflationary paradigms. It turns out that by  using only the power spectrum, it is not possible to distinguish between these two paradigms. This is due to the fact that the power spectrum at the end of both paradigms is found to be the same. However, the perturbations act differently
in both paradigms. In the standard slow-roll inflation, perturbations
freeze outside the horizon. However, in the matter bounce, perturbations grow
even outside the horizon. These different characteristic
behaviors in two different paradigms can be distinguished in the higher order correlation functions. For example, the non-gaussianity parameter in the squeezed limit is scale invariant, (i.e.,  \textit{consistency relation}) in the slow-roll inflation. However, in the matter bounce Universe, the consistency relation is violated and the non-gaussianity parameter becomes scale dependent. In this article, our objective is to find the effect of this phenomenon on CMB.

Due to the extreme difficulty in evaluating the scalar bispectrum in the bouncing scenario, in this article, we confine our study only to tensor perturbation. In Ref. \citep{Tahara2017}, authors studied the B-mode auto-bispectrum
in a generalized slow-roll inflation and showed that only the derivative coupling term between the scalar field and the Einstein tensor can boost the tensor bispectrum. In this work, however, we focus on the simplest matter bounce model  in the minimal Einstein frame \citep{Chowdhury2015}.  The objective is to study the same due to matter bounce and compare it with the standard slow-roll inflation.  

The paper is structured in the following manner. In the next section,
we obtain the relation of three point correlation function of spherical
harmonic coefficients in terms of three point function of primordial
perturbations. Analytic form of bispectrum due to matter bounce is
obtained in section \ref{subsec:Bispectrum-Inflation}. This, in section
\ref{sec:numerical-compute} is followed by a detailed discussion
of the numerical computation techniques for evaluating the
bispectrum. The computation is done with the help of transfer
functions obtained from CAMB\footnote{The source code can be downloaded from \href{https://camb.info}{https://camb.info}.}
software. In our analysis, we have used Planck 2018 cosmological parameters
\citep{Aghanim2018}. In section \ref{sec:Results}, we summarize
our results. Finally, in section \ref{sec:Conclu}, we conclude this
paper by comparing our results with the same in standard slow-roll
inflation scenario.

\section{\label{sec:Theory}CMB spectra}

CMB field can be characterized in terms of Stokes' parameters: $I$,
$Q$ and $U$, where $I$ represents temperature while $Q$ and $U$
linear polarization. It is known that $I$ (or $T$) is a spin 0 field
under a rotation of coordinate system in the tangent plane on the
surface of the sphere. So its spherical harmonic decomposition can
be expressed in the following manner: 
\begin{equation}
T\left(\theta,\phi\right)=\sum_{l=0}^{\infty}\sum_{m=-l}^{l}Y_{lm}T_{lm}.\label{eq:temp_decom}
\end{equation}

The polarization fields $Q$ and $U$, on the other hand transform
in a non-trivial manner. For example upon a right handed rotation about the normal
in the tangent plane by an angle $\psi$, the transformed fields satisfy
\citep{Zaldarriaga1996} 
\begin{equation}
\left(Q\pm iU\right)^{\prime}=e^{\mp2i\psi}\left(Q\pm iU\right).
\end{equation}

Due to this transformation property, $Q\pm iU$ are respectively called
as spin $\pm2$ fields. It would be desirable to obtain scalar fields
from $Q$ and $U$ as scalar fields are easier to
work with. A differential operator $\eth$ called `edth' \citep{Goldberg1966,Newman1966},
defined on the sphere's surface \citep{Zaldarriaga1996} can be used
on the combination $Q\pm iU$ to obtain scalars. These two scalars
are known as $E$ \& $B$ modes of CMB and are defined as follows
\begin{eqnarray}
E\left(\theta,\phi\right) &=&-\frac{1}{2}\left[\bar{\eth}^{2}\left(Q+iU\right)+\eth^{2}\left(Q-iU\right)\right],\\
B\left(\theta,\phi\right) &=& -\frac{i}{2}\left[\bar{\eth}^{2}\left(Q+iU\right)-\eth^{2}\left(Q-iU\right)\right].
\end{eqnarray}

Again since these fields are spin zero, we can perform a spherical
harmonic decomposition using standard spherical harmonics $Y_{lm}$'s:
\begin{equation}
X\left(\theta,\phi\right)=\sum_{l=2}^{\infty}\sum_{m=-l}^{l}X_{lm}Y_{lm}\left(\theta,\phi\right),\ X=E,B.\label{eq:pola_decomp}
\end{equation}

Please note that in Eq. (\ref{eq:temp_decom}), sum over $l$
starts from $l=0$ where as in Eq. (\ref{eq:pola_decomp}), from $l=2$.
This is due to the properties of $Q$ and $U$ fields, as they are
described in terms of spin 2 spherical harmonics \citep{Tevain1985,Tevain1986}.
The $T$, $E$ and $B$ mode spherical harmonic coefficients, given
respectively in Eqs. (\ref{eq:temp_decom}) and (\ref{eq:pola_decomp})
can be related to primordial perturbations in the following manner
\citep{Shiraishi2012} 
\begin{equation}
X_{lm}^{(Z)}=4\pi\left(-i\right)^{l}\sum_{s}\int\frac{d^{3}\mathbf{k}}{\left(2\pi\right)^{3}}\ _{-s}Y_{lm}^{*}\left(\Omega_{\mathbf{k}}\right)\left(\text{sgn}\left(s\right)\right){}^{s+x}\xi^{(s)}\left(\mathbf{k}\right)T_{X,l}^{(Z)}\left(k\right)\label{eq:X_Mode_Har_Coeff}
\end{equation}

We next explain the meaning of each term in the above expression.
First, $\xi^{(s)}$ is the primordial perturbation corresponding to
a given helicity $s$ that takes the following values:

\[
s=\begin{cases}
0 & Z=S\\
\pm1 & Z=V\\
\pm2 & Z=T.
\end{cases}
\]

$Z$ denotes the nature of perturbation (Scalar, Vector or Tensor).
The index $x$ depends upon the field being considered, takes the
following values 
\[
x=\begin{cases}
0 & X=T,E\\
1 & X=B.
\end{cases}
\]

$T_{X,l}^{(Z)}\left(k\right)$ is the transfer function for a given
field $X$ (which can be $T$, $B$ or $E$) and the nature of perturbation
$Z$. The symbol $\text{sgn}\left(x\right)$ stands for the {\it signum}
function which is defined as 
\[
\text{sgn}\left(x\right)=\begin{cases}
1 & x>0\\
0 & x=0\\
-1 & x<0,
\end{cases}
\]
and $_{s}Y_{lm}$ are spin weighted spherical harmonics.

Therefore, if we can evaluate $\xi^{(s)}({\bf k})$ which depends on the theory, we can compute $X_{lm}^{(Z)}$ using Eq. (\ref{eq:X_Mode_Har_Coeff}) which in turn helps to evaluate the CMB spectra. 

The Horndeski theory \citep{Horndeski1974,Deffayet2009,Deffayet2010a,Deffayet2010b} is the most general scalar-tensor theory in four dimensions. The specialty of this theory is that the equations of motion lead to second order differential equations and therefore the theory is free from Ostrogradsky instabilities \citep{Ostro1850}. The action is given as
\begin{eqnarray}
S_g = \int d^4x\, \sqrt{-g}\,\left(\mathcal{L}_2 + \mathcal{L}_3 + \mathcal{L}_4 + \mathcal{L}_5 \right),
\end{eqnarray}
where the $\mathcal{L}_{i}$'s are defined in the following manner
\begin{eqnarray}
\mathcal{L}_{2} &=& K\left(\phi,X\right), \\
\mathcal{L}_{3} &=& -G_{3}\left(\phi,X\right)\square\phi,\\
\mathcal{L}_{4} &=& G_{4}\left(\phi,X\right)R+G_{4X}\left[\left(\square\phi\right)^{2}-\left(\nabla_{\mu}\nabla_{\nu}\phi\right)^{2}\right], \\
\mathcal{L}_{5} &=& G_{5}\left(\phi,X\right)G_{\mu\nu}\nabla^{\mu}\nabla^{\nu}\phi -\frac{1}{6}G_{5X}\Big[\left(\square\phi\right)^{3}-3\square\phi\left(\nabla_{\mu}\nabla_{\nu}\phi\right)^{2}+2\left(\nabla_{\mu}\nabla_{\nu}\phi\right)^{3}\Big]. 
\end{eqnarray}

\noindent $X \equiv -1/2\,g^{\mu \nu}\, \nabla_{\mu}  \phi \nabla_\nu \phi$ is the kinetic term, $R$ is the Ricci scalar and $G_{\mu\nu}$ the
Einstein tensor. $K(\phi, X), G_3 (\phi, X), G_4 (\phi, X)$ and $G_5(\phi, X)$ are functions of $\phi$ and $X$ and the subscripts $\phi$ and  $X$ denote the partial derivatives with respect to the corresponding variables. In the FRW background, the metric components with the tensor perturbations in cosmic time $t$ can be written as
\begin{equation}
g_{00}=-1,\ \ \ \ g_{0i}=0,\ \ \ \ g_{ij}=a^{2}(t)(e^{h})_{ij},
\end{equation}
where
\begin{equation}
(e^{h})_{ij}=\delta_{ij}+h_{ij}+\frac{1}{2}h_{ik}h_{kj}+\ldots,
\end{equation}
$a(t)$ is the scale factor and $h_{i j}$ is the tensor perturbation. Using this, the quadratic and cubic actions for $h_{ij}$ can be written
as
\begin{eqnarray}
S^{(2)} &=& \frac{1}{8}\int d^{4}x\, a^{3}\left[\mathcal{G}_{T}\dot{h}_{ij}^{2}-\frac{\mathcal{F}_{T}}{a^{2}}\left(\partial_{k}h_{ij}\right)^{2}\right],\label{eq:qua_act}\\
S^{(3)} &=& \int d^{4}x\,a^{3}\left[\frac{\mathcal{F}_{T}}{4a^{2}}\left(h_{ik}h_{jl}-\frac{1}{2}h_{ij}h_{kl}\right)\partial_{k}\partial_{l}h_{ij}\right. +\left.\frac{X\dot{\phi}G_{5X}}{12}\dot{h}_{ij}\dot{h}_{jk}\dot{h}_{kl}\right]\label{eq:cub_act},
\end{eqnarray}
where
\begin{eqnarray}
\mathcal{F}_{T} &= &2\left[G_{4}-X\left(\ddot{\phi}G_{5X}+G_{5\phi}\right)\right],\\
\mathcal{G}_{T} &=& 2\left[G_{4}-2XG_{4X}-X\left(H\dot{\phi}G_{5X}-G_{5\phi}\right)\right].
\end{eqnarray}

\noindent $\xi^{(s)}({\bf k})$ for the tensor perturbation is defined as
\[
\xi^{(s)}\left(\mathbf{k}\right)=h_{jk}\left(\mathbf{k}\right)e_{jk}^{*(s)}\left(\mathbf{k}\right),
\]
where $e_{ij}^{(s)}$ is the polarization tensor and $e^{(s)}_{i j}({\bf k})\, e^{* (s)}_{i j}({\bf k}) = 2.$

\subsection{Power Spectrum}

Using the quadratic action (\ref{eq:qua_act}), we can solve for the mode function $\xi^{(s)}({\bf k})$ and evaluate the two point correlation function as

\begin{equation}
\left\langle \xi^{(s)}\left(\mathbf{k}\right)\xi^{*(s^{\prime})}\left(\mathbf{k}^{\prime}\right)\right\rangle =\left(2\pi\right)^{3}\delta^{(3)}\left(\mathbf{k}-\mathbf{k}^{\prime}\right)\delta_{ss^{\prime}}\, P_{T}(k),
\end{equation}
here $P_{T}(k)$ is the primordial tensor power spectrum which can be further written as
\begin{equation}
P_T(k) = \frac{\pi^{2}}{k^{3}}\mathcal{P}_{h}(k).\label{eq:2_pow_spec}
\end{equation}
 
\noindent $\mathcal{P}_{h}(k)$ is referred to as the dimensionless tensor power spectrum. In case of slow-roll inflation, it takes the form
\begin{equation}
\mathcal{P}^{\rm inf}_{h}(k)=\frac{2H^{2}}{\pi^{2}}\sqrt{\frac{\mathcal{G}_{T}}{\mathcal{F}_{T}^{3}}}.
\end{equation}

\noindent However, in the case of matter bounce with the scale factor $a(\eta)  = a_0 \,(1 + k_0^2 \eta^2)$, it is difficult to obtain the general solution of the mode function as the time dependencies of the functions  $G_4 (\phi, X)$ and $G_5(\phi, X)$ are not known. In the case of simplest minimal coupling with $G_4 (\phi, X) = 1/2$ and $G_5(\phi, X) = 0$, the solution and therefore the primordial power spectrum $\mathcal{P}^{\rm mb}_{h}(k)$  is known \cite{Chowdhury2015}. 

Once we obtain the primordial power spectrum, using Eq. (\ref{eq:X_Mode_Har_Coeff}),
we can calculate the two point function of B-mode harmonic coefficients
as \citep{Shiraishi2012}
\begin{align}
\left\langle B_{l_{1}m_{1}}^{(T)}B_{l_{2}m_{2}}^{(T)*}\right\rangle  & =\delta_{l_{1}l_{2}}\delta_{m_{1}m_{2}}C_{l_{1}}^{BB},\label{eq:ang_pow_spec}
\end{align}
where the B-mode angular power spectrum $C_{l}^{BB}$ is defined to
be
\begin{equation}
C_{l}^{BB}=\frac{2}{\pi}\int k^{2}dkP_{T}(k)T_{B,l}^{(T)}(k)T_{B,l}^{(T)}(k)
\end{equation}
with $P_T(k)$ defined in Eq. (\ref{eq:2_pow_spec}).

\begin{figure}[t]
\centering
\includegraphics[scale=0.25]{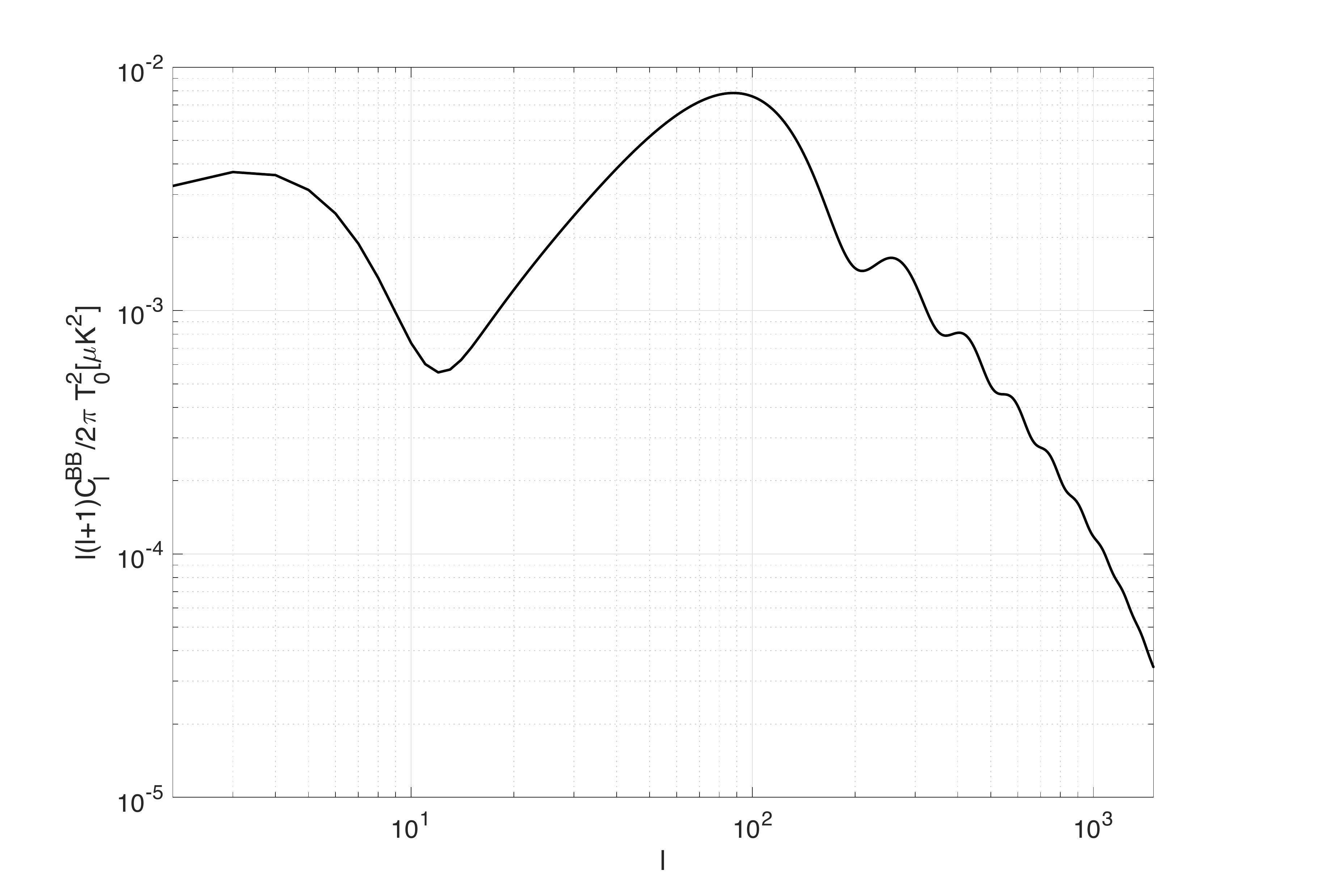}
\caption{\label{fig:B-Mode-power}B-mode power spectrum due to inflation as well as for matter bounce.}
\end{figure}

In case of slow-roll inflation as well as for the simplest matter bounce scenario prescribed in Ref. \citep{Chowdhury2015}, B-mode power spectrum turns out to be identical and is shown in Figure \ref{fig:B-Mode-power}. Therefore, in order to see the differences in these two cases, we need to go beyond the power spectrum and evaluate the B-mode auto Bispectrum for these two cases.

\subsection{Bispectrum}

Since, in the case of B-mode power spectrum, slow-roll inflation as well as matter bounce behave identically, therefore, by studying CMB power spectrum,
it is not possible to distinguish between the two paradigms. However,
there are characteristic differences between the perturbations in
two different paradigms. As mentioned before, perturbations freeze outside
the horizon during inflation, whereas, they grow in the matter bounce Universe.
This behavior is encoded in the higher order correlation functions.
Due to this reason, in this work, we concentrate on the three point
correlation function, i.e., the bispectrum. In this section, we give the explicit form of the bispectrum due to the simplest matter bounce scenario with $G_4 = 1/2, G_5 = 0$ \citep{Chowdhury2015} and compare the same with that due to inflation
given in \citep{Tahara2017}.

B-mode auto-bispectrum $B_{l_{1}l_{2}l_{3}}$ in terms of the three
point correlations of the B-mode harmonic coefficients is defined
as 
\begin{equation}
B_{l_{1}l_{2}l_{3}}=\sum_{m_{i}}\begin{pmatrix}l_{1} & l_{2} & l_{3}\\
m_{1} & m_{2} & m_{3}
\end{pmatrix}\left\langle \prod_{i=1}^{3}B_{l_{i}m_{i}}\right\rangle .\label{eq:crude_B_Mode_Bispec}
\end{equation}

In this expression, the quantity inside the parenthesis is the Wigner
3j symbol which is related to the Clebsch Gordon coefficients which
appear while adding angular momenta. Using Eq. (\ref{eq:X_Mode_Har_Coeff}), $B_{l_{1}l_{2}l_{3}}$ in  Eq. (\ref{eq:crude_B_Mode_Bispec})
can also be written as

\begin{equation}
B_{l_{1}l_{2}l_{3}}=\sum_{s_{i}=\pm2}B_{l_{1}l_{2}l_{3}}^{\left(s_{1}s_{2}s_{3}\right)},\label{eq:b_mode_auto_bispec}
\end{equation}
where, 
\begin{eqnarray}
B_{l_{1}l_{2}l_{3}}^{\left(s_{1}s_{2}s_{3}\right)}&=&\sum_{m_{i}}\begin{pmatrix}l_{1} & l_{2} & l_{3}\\
m_{1} & m_{2} & m_{3}
\end{pmatrix}\prod_{j=1}^{3}\left[4\pi\left(-i\right)^{j}\int\frac{d^{3}\mathbf{k}}{\left(2\pi\right)^{3}}(\text{sgn}(s_{j})){}^{s_{j}+1}\ _{-s_{j}}Y_{l_{j}m_{j}}^{*}(\Omega_{\mathbf{k}_{j}})T_{l_{j}}(k_{j})\right]\nonumber\\
&&\times\left\langle \xi^{\left(s_{1}\right)}\left(\mathbf{k}_{1}\right)\xi^{\left(s_{2}\right)}\left(\mathbf{k}_{2}\right)\xi^{\left(s_{3}\right)}\left(\mathbf{k}_{3}\right)\right\rangle .\label{eq:genera_form_bispec}
\end{eqnarray}

Finally, using Eq. \eqref{eq:cub_act}, the three point correlation of the primordial perturbations $\xi^{\left(s_{i}\right)}$ in the absence of $G_{5X}$ can be written as 
\begin{eqnarray}
\left\langle \xi^{\left(s_{1}\right)}\left(\mathbf{k}_{1}\right)\xi^{\left(s_{2}\right)}\left(\mathbf{k}_{2}\right)\xi^{\left(s_{3}\right)}\left(\mathbf{k}_{3}\right)\right\rangle  =\left(2\pi\right)^{7}f_{\mathbf{k}_{1}\mathbf{k}_{2}\mathbf{k}_{3}}^{s_{1}s_{2}s_{3}}
  \times\delta^{3}\left(\mathbf{k}_{1}+\mathbf{k}_{2}+\mathbf{k}_{3}\right), \label{eq:primor_three_heli_s}
\end{eqnarray}

where the function $f_{\mathbf{k}_{1}\mathbf{k}_{2}\mathbf{k}_{3}}^{s_{1}s_{2}s_{3}}$
can be written as 
\begin{equation}
 f_{\mathbf{k}_{1}\mathbf{k}_{2}\mathbf{k}_{3}}^{s_{1}s_{2}s_{3}}=\mathcal{G}\left(k_{1},k_{2},k_{3}\right)\times\left[\left(e_{im}^{s_{2}}e_{lj}^{s_{3}}-\frac{1}{2}e_{ml}^{s_{2}}e_{ij}^{s_{3}}\right)^{*}e_{ij}^{s_{1}*}k_{1m}k_{1l}+5\text{ Perms}\right].\label{eq:f_func}
\end{equation}

The angular dependence comes with the polarization tensor $e_{ij}^{s_{i}}\equiv e_{ij}^{s_{i}}(\Omega_{\mathbf{k}_{i}})$
which is expressible in terms of spin $s$ weighted spherical harmonics.
This quantity is calculated from the cubic order action given in Eq.
\eqref{eq:cub_act}. The form of the function $\mathcal{G}\left(k_{1},k_{2},k_{3}\right)$
in Eq. (\ref{eq:f_func}) depends upon the model being employed. In case of slow-roll inflation, $\mathcal{G}\left(k_{1},k_{2},k_{3}\right)$ is given in Ref. \citep{Gao2011}  whereas, in our case, for simple matter bounce, the expression is obtained in Ref. \citep{Raveendran2017}. 

\section{\label{subsec:Bispectrum-Inflation}Bispectrum due to Matter Bounce}

On account of the properties of Wigner 9j symbols in Eq. (\ref{eq:genera_form_bispec}),
that appear after performing angular integrations, it turns out that
only for $\sum l_{i}=\text{odd}$, the B-mode auto-bispectrum in Eq.
(\ref{eq:b_mode_auto_bispec}) is nonzero. Further, the bispectrum
is  nonzero only  when all $l_{i}$'s are different.

In this work, we calculate bispectrum for $\sum l_{i}=33$. This implies
114 possibilities in total for the triple $\left(l_{1},l_{2},l_{3}\right)$.
However, due to symmetry, we need to evaluate only one of the six
permutations and rest others can be evaluated by noting the sign of
permutation.

We obtain the analytical form of $B_{l_{1}l_{2}l_{3}}$ in Eq. \eqref{eq:b_mode_auto_bispec} due to simple matter bounce after substituting
the simplified form of the three point correlations of primordial
perturbations from Eq. (\ref{eq:g_func_bounce}) in Eq. (\ref{eq:genera_form_bispec}).
Since we are considering only a specific regime (as per the discussions
done in Appendix \ref{sec:Simplification-of-Bispectrum-Matter}) which
contributes the most, we get the following approximate expression:

\begin{align}
B_{l_{1}l_{2}l_{3}} & \approx C\sum_{L_{i},l_{i}^{\prime}}\int_{0}^{\infty}x^{2}dx\left[\prod_{j=1}^{3}\int_{0}^{\infty}k_{j}^{2}dk_{j}T_{l_{j}}\left(k_{j}\right)j_{L_{j}}\left(xk_{j}\right)\right]\left[k_{1}^{2}\left(3\sqrt{3}\delta_{l_{1}^{\prime},2}-\delta_{l_{1}^{\prime},4}\right)\delta_{l_{2}^{\prime},2}\delta_{l_{3}^{\prime},2}+\text{ 2 perms}\right]\nonumber \\
 & \times\left(i^{\sum L_{j}-l_{j}}\right)I_{L_{1},L_{2},L_{3}}^{0,0,0}I_{l_{1},L_{1},l_{1}^{\prime}}^{-2,0,2}I_{l_{2},L_{2},l_{2}^{\prime}}^{-2,0,2}I_{l_{3},L_{3},l_{3}^{\prime}}^{-2,0,2}\begin{Bmatrix}l_{1}^{\prime} & l_{2}^{\prime} & l_{3}^{\prime}\\
L_{1} & L_{2} & L_{3}\\
l_{1} & l_{2} & l_{3}
\end{Bmatrix}\left[\frac{1}{\left(k_{1}k_{2}\right)^{3}}+\frac{1}{\left(k_{2}k_{3}\right)^{3}}+\frac{1}{\left(k_{3}k_{1}\right)^{3}}\right],\label{eq:bispec_simplified}
\end{align}

here the symbol $I_{L_{1}L_{2}L_{3}}^{s_{1}s_{2}s_{3}}$ in terms
of Wigner 3j symbols is defined as 
\[
I_{L_{1}L_{2}L_{3}}^{s_{1}s_{2}s_{3}}=\sqrt{\frac{\prod_{i=1}^{3}\left(2L_{i}+1\right)}{4\pi}}\begin{pmatrix}L_{1} & L_{2} & L_{3}\\
s_{1} & s_{2} & s_{3}
\end{pmatrix}.
\]
The explicit expression of $C$ in Eq.  (\ref{eq:bispec_simplified}) is given in Eq. (\ref{eq:constant}).
The simplification part of the three point function is discussed in
Appendix \ref{sec:Simplification-of-Bispectrum-Matter}.

Presence of the Wigner symbols in the above expression assigns specific
values to $L_{i}$'s for given $l_{i}$'s. For example, consider the
Wigner 9j symbol. 
\[
\begin{Bmatrix}l_{1}^{\prime} & l_{2}^{\prime} & l_{3}^{\prime}\\
L_{1} & L_{2} & L_{3}\\
l_{1} & l_{2} & l_{3}
\end{Bmatrix}.
\]
This symbol can be written in terms of a sum over Wigner 3j symbols
\citep{Varshalovich1988}. The Wigner symbols imply the following
conditions 
\begin{equation}
\left|l_{1}-l_{2}\right|\le l_{3}\le l_{1}+l_{2},\ \left|l_{i}-l_{i}^{\prime}\right|\le L_{i}\le l_{i}+l_{i}^{\prime}.\label{eq:ell_conds}
\end{equation}
This means that we need to perform integrations over only certain
combinations of $l$ and $L$. This turns out to be very helpful in
devising numerical integration technique which is discussed in the
next section. The results of numerical computation are given in section
\ref{sec:Results} where we do the comparison of the bispectrum between
the two paradigms.

\section{\label{sec:numerical-compute}Numerical Computation of Bispectrum}

In this section, we discuss the numerical techniques for the evaluation
of the bispectra due to both inflation and the bounce. The details
of the results, these techniques are based on, can be found in Appendix
(\ref{sec:Numerical-Strategies}). The steps for calculating the bispectrum
are as follows: 

\begin{enumerate}
\item First of all, the values of $l_{i}$ and $L_{i}$'s are chosen such
that they satisfy all the requirements of the Wigner 3j and 9j symbols
present in Eq. (\ref{eq:bispec_simplified}). Thus for our case, in
addition to the conditions in Eq. (\ref{eq:ell_conds}), we must also
have $\sum L_{i}=\text{even}$ because of the presence\footnote{Weisstein, Eric W. ``Wigner 3j-Symbol.'' From Mathworld -- A Wolfram
Web Resource \href{http://mathworld.wolfram.com/Wigner3j-Symbol.html}{http://mathworld.wolfram.com/Wigner3j-Symbol.html}} of $I_{L_{1}L_{2}L_{3}}^{000}$. 
\item As can be seen from Eq. (\ref{eq:bispec_simplified}) that we need to
perform four integrations. One is over $x$ and other three are on
$k_{i}$'s. 
\item The procedure for integrating over $k$ variable was based on the
presence/absence of the term $k_{t}=\sum k_{i}$ in the denominator.
If it is absent, we will call bispectrum as \textit{separable} and \textit{inseparable}
otherwise. Thus we take two cases 
\begin{itemize}
\item[A.] For separable form we evaluate the integral 
\begin{equation}
I_{x}\left(n,l,L\right)=\int k^{n}dkT_{l}\left(k\right)j_{L}\left(kx\right)\label{eq:the_num_integ}
\end{equation}
for different values of $x$. 
\item[B.] For an inseparable form, we first take Laplace's transform and introduce
another variable $y$ \citep{Tahara2017} and evaluate the following
integral for different $x$ and $y$'s 
\begin{equation}
I_{x,y}\left(n,l,L\right)=\int k^{n}dkT_{l}\left(k\right)j_{L}\left(kx\right)e^{-ky}.\label{eq:the_num_integ-Lapalce}
\end{equation}
\end{itemize}
\item These integrals are then stored in an array for the relevant values
of $n$, $l$ and $L$ for given $x$ (and $y$). We here emphasize
that this integral evaluation part can easily be parallel implemented. 
\item In order to retrieve integrals while doing integration over $x$ (and
$y$) for different values of $n$, $l$ and $L$, we use the binary
search method. We explain the method using an example later in this
section. The method relies on two results derived in Appendix \ref{subsec:Binary-Search}. 
\item We also need to compute Wigner 3j and 9j symbols. We did this computation
using two methods which we also elaborate on later in this section. 
\item The condition $\sum l_{i}=33$ along with Wigner 9j symbol gave 114
possibilities on the triple $(l_{1},l_{2},l_{3})$. The bispectrum
will be same modulo a phase sign for different permutations of $(l_{1},l_{2},l_{3})$.
This reduces the computation to 1/6. 
\end{enumerate}
All of the steps just discussed can be summarized in the flow chart
shown in Figure \ref{fig:Flow-chart} below.

\begin{figure}[t]
\begin{centering}
\includegraphics[scale=0.35]{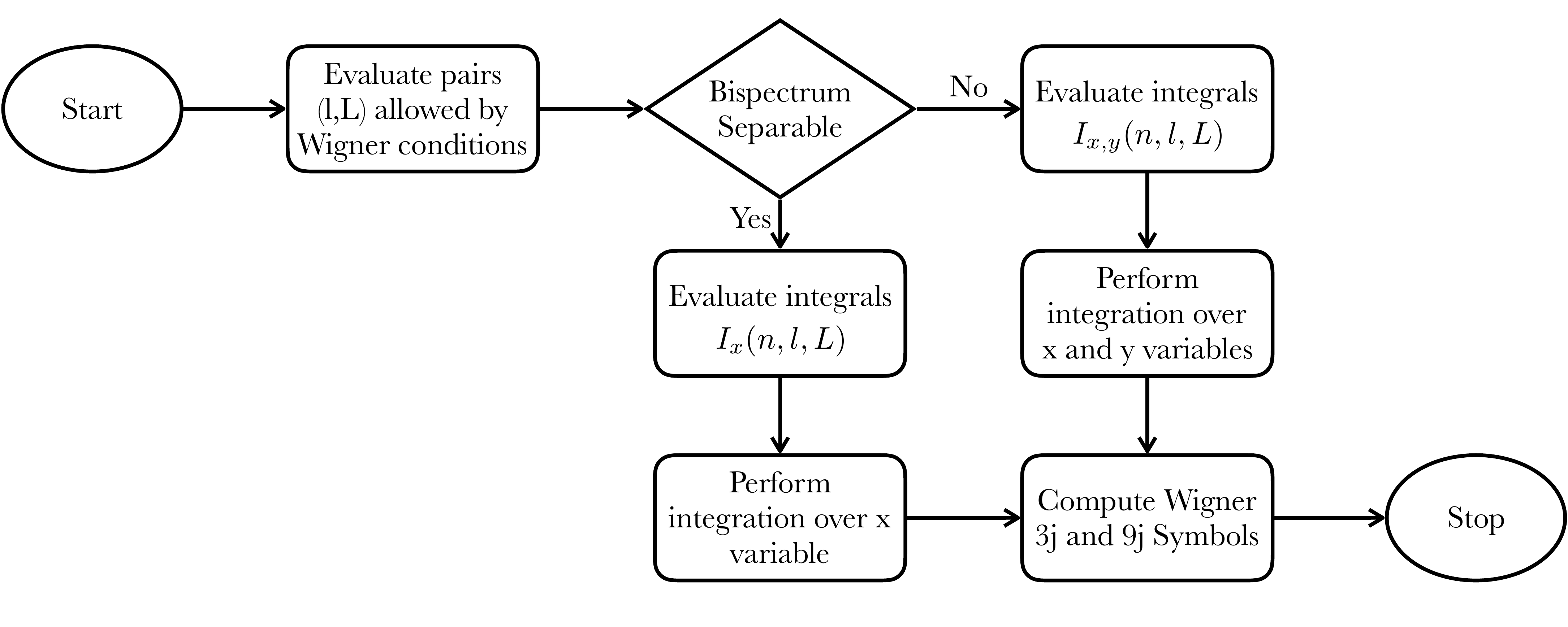} 
\par\end{centering}
\caption{\label{fig:Flow-chart}Flow chart of the algorithm for numerical calculation
of the bispectrum. This very algorithm can be used for computing more complicated forms of bispectra.}
\end{figure}
\subsection{Example of Binary Search Method}

Next, we explain the $n$ tuple binary search method with an example.
Suppose we have the following 3 tuple array 
\[
\{(1,1,3),(1,2,4),(1,3,-3),(2,0,5),(3,4,-2)\}.
\]

Let us assume that we want to find out the location of the tuple $(3,4,-2)$.
One obvious way would be to implement a linear search and to find
locations of integers 3, 4 and $-2$, thereby finding the location
of the tuple. A more efficient way would be to convert all of these
tuples into an integer and search for just that integer using Theorem 1 of section (\ref{subsec:Binary-Search}). Also notice that the tuple array is in the
lexicographic order in the sense described in Appendix \ref{sec:Numerical-Strategies}.

Now we apply Theorem 1 of section (\ref{subsec:Binary-Search}), according to which
the \textit{base of representation} would be $M\ge9$. For $M=9$,
the map would convert these tuples into the following 
\[
\{93,103,105,167,277\}.
\]

Here we can see that the integers appear in ascending order as
per Theorem 2 of section (\ref{subsec:Binary-Search}). After this, we can implement
a binary search so that in place of searching the tuple $(3,4,-2)$,
we just search the integer 277. This method can be used to store the
integral $I_{x}(n,l,L)$ corresponding to say $(n,l,L)=(-2,1,3)$
and to later retrieve it while performing $x$ integration.

We must emphasize that the binary search method for tuples can be applied
in all places where the location of a tuple is sought after storage.
The method becomes more and more advantageous as the length of the
tuple becomes larger and larger.

\subsection{Wigner Symbols' Computation}

We now provide details about the Wigner 3j calculation. Since 9j symbols
can be expressed in terms of 3j symbols, the discussion of 3j would
suffice. We employed two methods of computation.

The first method is the Rasch Algorithm \citep{Rasch2004}. This is
based on the Regge symmetries of the Wigner 3j symbols. The Rasch
method employs 36 out of 72 symmetries\footnote{Source code: \href{https://github.com/ramanujakothari/RaschAlgo}{https://github.com/ramanujakothari/RaschAlgo}}.
Thus by computing and storing one symbol, 36 other symbols are determined
as well. This considerably reduces the computation time. For the evaluation
of the symbols, we use \texttt{fgsl}\footnote{Link \href{https://www.lrz.de/services/software/mathematik/gsl/fortran/}{https://www.lrz.de/services/software/mathematik/gsl/fortran/}}
which is a Fortran version of \texttt{gsl}\footnote{Link \href{https://www.gnu.org/software/gsl/}{https://www.gnu.org/software/gsl/}}.
But it turned out that \texttt{fgsl} gives \texttt{segmentation fault}
at larger values of l. So we use \texttt{wigxjpf} for the bispectrum
evaluation at larger values of $l$ while calculating SNR.

\texttt{Wigxjpf} software \citep{Johansson2016} calculates these
symbols using prime number factorization. 

\begin{figure}[t!]
\begin{centering}
\subfloat[Slow-roll Inflation]{\begin{centering}
\includegraphics[scale=0.44]{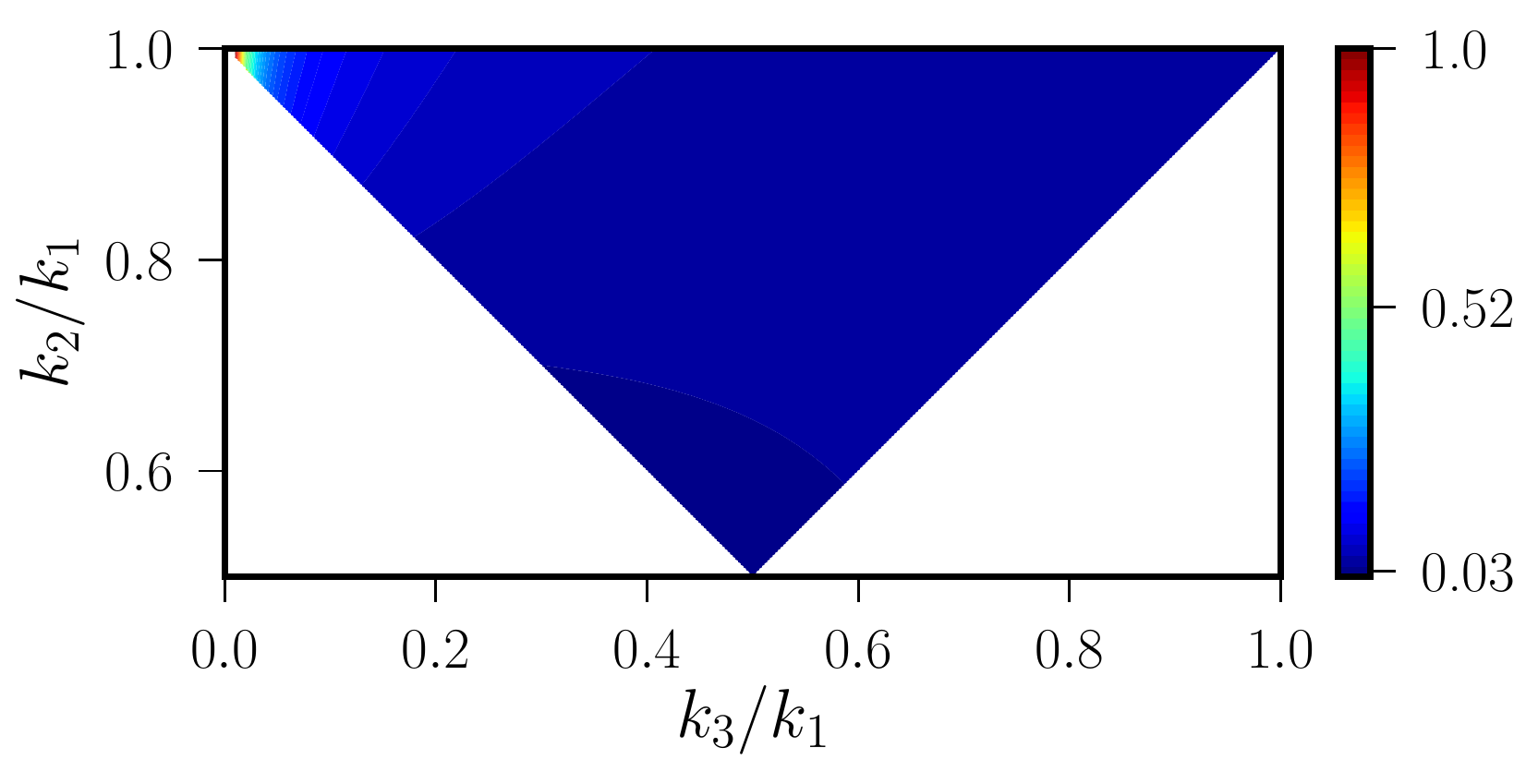} 
\par\end{centering}
}\subfloat[Bounce]{\begin{centering}
\includegraphics[scale=0.44]{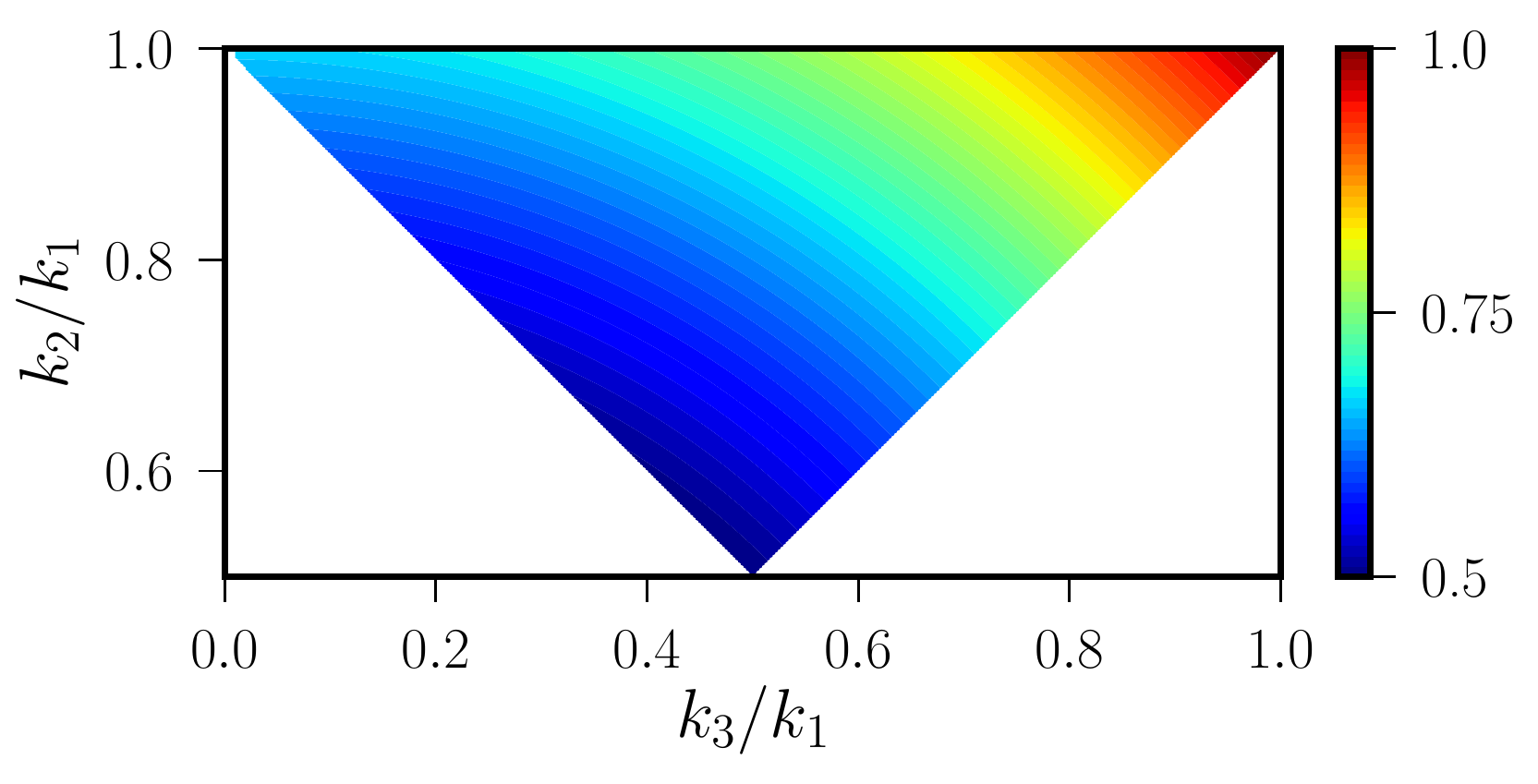} 
\par\end{centering}
}
\par\end{centering}
\caption{\label{fig:Comparison-of-the-Shape}Comparison of the shape function
normalized to unity.}
\end{figure}

\begin{figure}[ht!]
\subfloat[Inflation]{\begin{centering}
\includegraphics[scale=0.3]{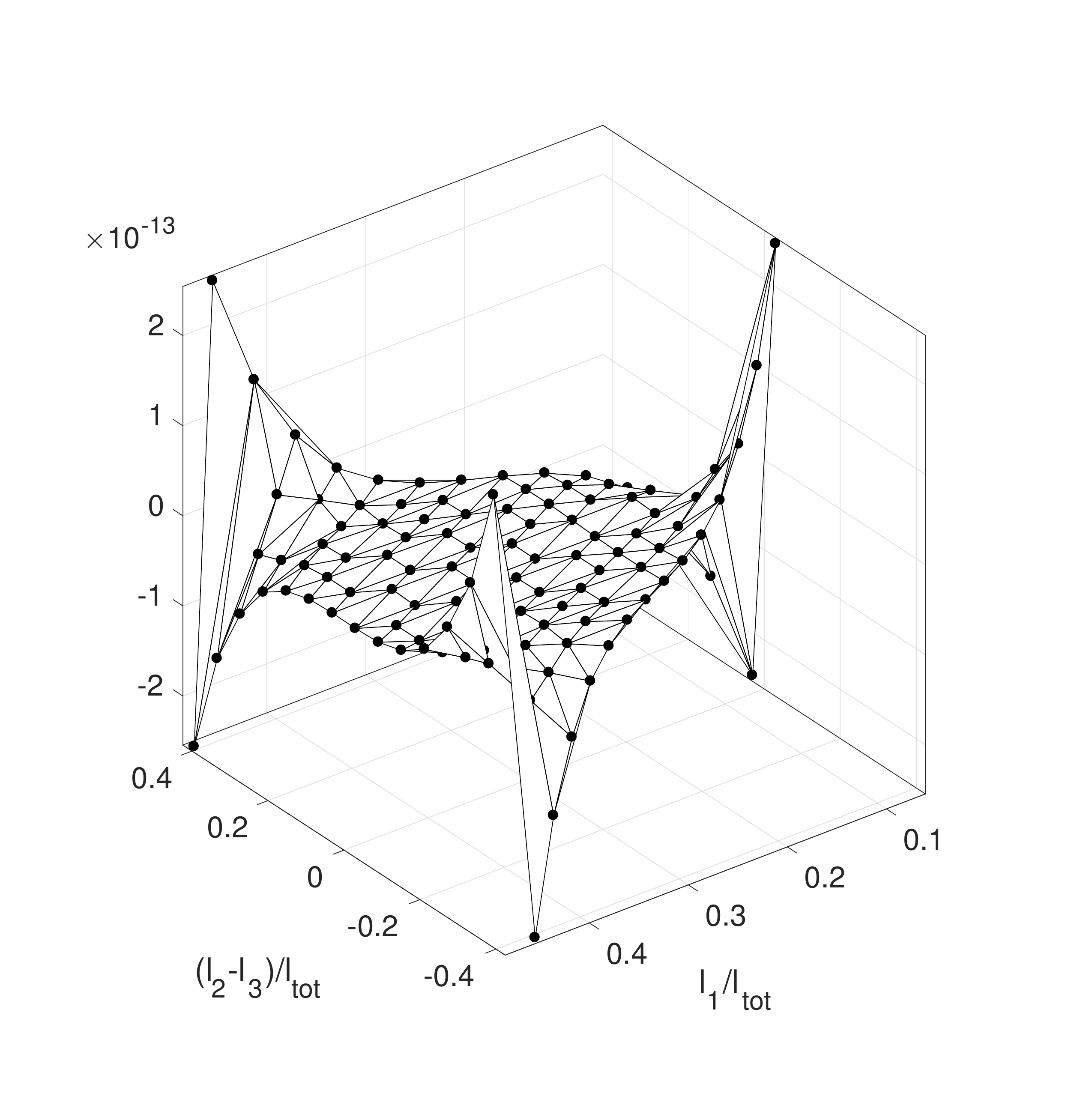} 
\par\end{centering}
}\subfloat[Bounce]{\begin{centering}
\includegraphics[scale=0.3]{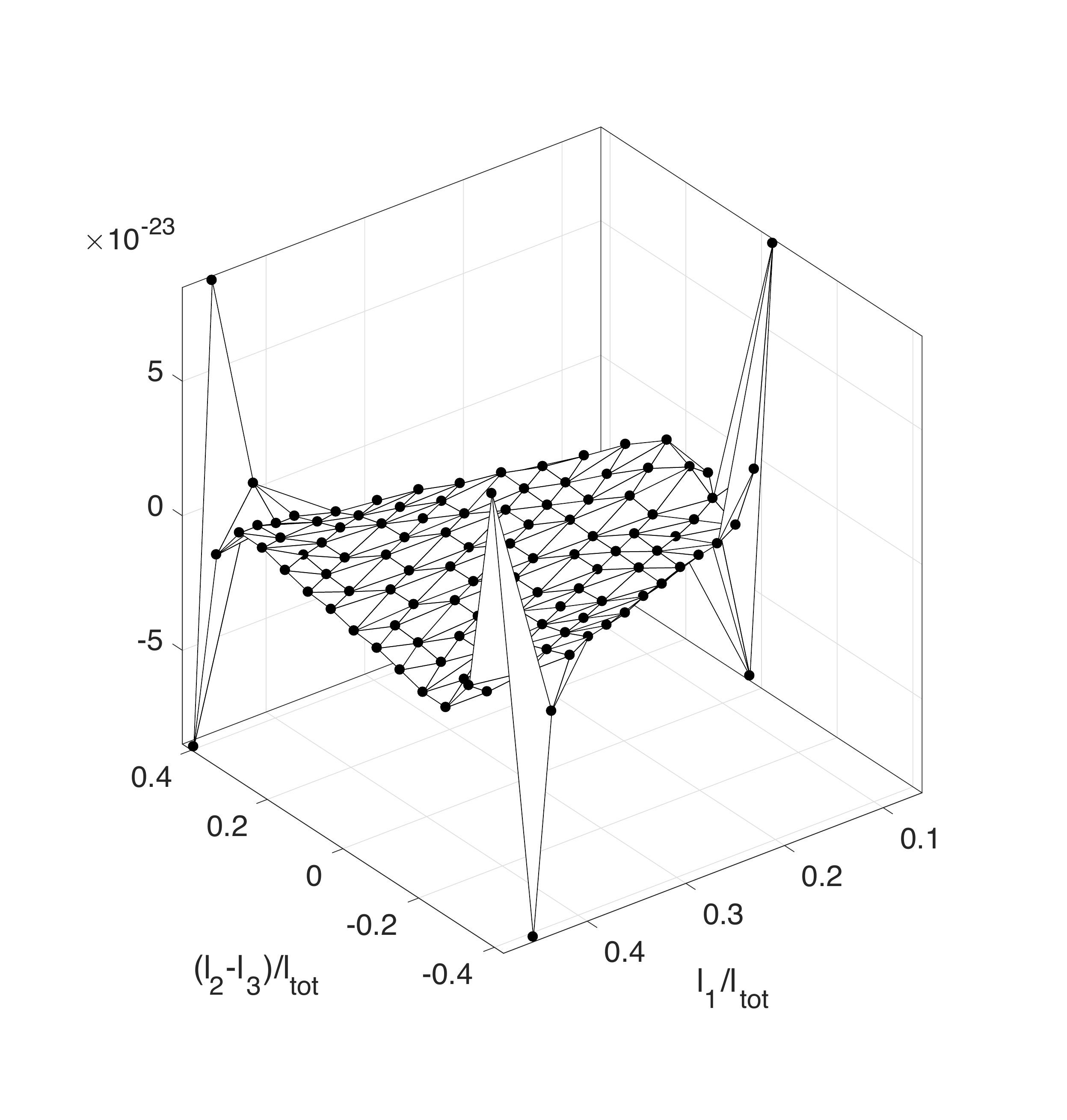} 
\par\end{centering}
}\caption{\label{fig:Comparison-of-the-Bispectra}Comparison of the Bispectra
plots for inflation and bounce cases. The bispectra values are obtained
after taking tensor to scalar ratio $r=0.01$. In both cases, we have
divided the bispectrum by the square of the tensor amplitude i.e.,
$\mathcal{P}_{T}^{2}$.}
\end{figure}

\section{\label{sec:Results}Results}

In this section, we give all the results of our numerical computation.
First of all, we compare the shape function or the non-gaussianity
parameter $h_{\text{NL}}$ for inflation with that of matter bounce.
This can be found in Figure \ref{fig:Comparison-of-the-Shape}. We
see that the two bispectra are different for all three limits - squeezed,
folded and equilateral.

The bispectra results for the two cases can be found in Figure \ref{fig:Comparison-of-the-Bispectra}.
These plots have been generated by considering the tensor-to-scalar
ratio\footnote{This was done for comparing our results with those given in \citep{Tahara2017}
.} $r=0.01$. We see that the peaks of the bispectrum in both cases
appear at the permutations of $(l_{1},l_{2},l_{3})=(2,15,16)$. But
the magnitude is very much different. In the case of inflation, the highest
value of the bispectrum is found to be $\sim6\times10^{-11}$ while
in bounce it is about $\sim5\times10^{-23}$. This feature is clearly
different from the angular power spectrum where the values from both
paradigms were the same. Thus the bispectrum is indeed able to distinguish
between the two paradigms. 

SNR is given by the following expression \citep{Tahara2017} 
\begin{equation}
\sqrt{\sum_{l_{i}\le l_{\text{max}}}\frac{B_{l_{1}l_{2}l_{3}}^{2}}{6C_{l_{1}}C_{l_{2}}C_{l_{3}}}}.
\end{equation}
here $l_{\text{max}}$ is the value of maximum multipole $l$ used
in the computation.

\begin{figure}[ht!]
\begin{centering}
\includegraphics[scale=0.25]{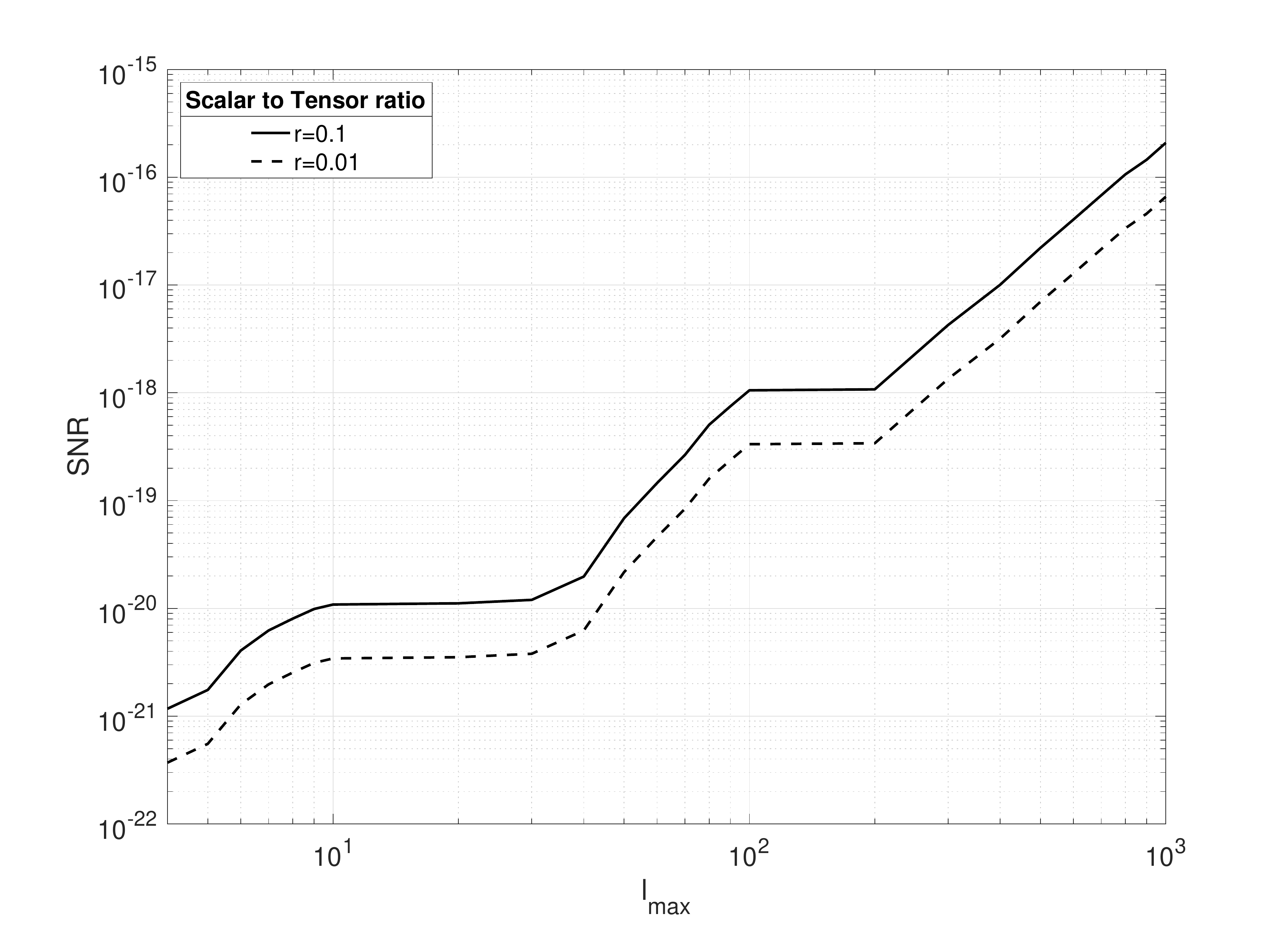} 
\par\end{centering}
\caption{\label{fig:Signal-to-Noise}SNR for the bounce model
for tensor-to-scalar ratio $r\in\{0.1,0.01\}$. }
\end{figure}

This expression for the SNR is valid for all triples $\left(l_{1},l_{2},l_{3}\right)$
which satisfy the conditions enforced by the Wigner 9j symbol, as
was discussed above. The bispectrum for all permutations of say $\left(2,3,5\right)$
will be the same, apart from a phase factor. Thus the expression can
be simplified by summing over only unique triples and thus can be
written as 
\begin{eqnarray}
	\sqrt{\sum_{l_{i}^{\text{unq}}\le l_{\text{max}}}\frac{B_{l_{1}l_{2}l_{3}}^{2}}{C_{l_{1}}C_{l_{2}}C_{l_{3}}}},
\end{eqnarray}
where $l_{i}^{\text{unq}}$ depicts the fact that the sum is over
unique triples. The SNR plot as a function of $l_{\text{max}}$ for
tensor to scalar ratio $r\in\left\{ 0.1,0.01\right\} $ is shown in
Figure \ref{fig:Signal-to-Noise}. For $r=0.01$ and $l_{\text{max}}=100$,
the SNR due to standard inflation turns out to be $\sim10^{-5}$,
whereas for bounce it is $\sim10^{-18}$. Thus the SNR due to bounce
is $\sim10^{-13}$ times smaller than due to inflation, thereby making
the signal due to bounce very difficult to detect. So a future detection of tensor mode bispectrum will be helpful in  ruling out matter bounce model.

\section{Conclusion}

\label{sec:Conclu} In this paper, we have analytically calculated
the B-mode auto-bispectrum due to matter bounce. We also have computed
the bispectrum numerically and compared it with that generated due
to the standard inflationary paradigm. The numerical computation of bispectrum
involves integration over 4 (or 5) integrals. Since the bispectrum $B_{l_{1}l_{2}l_{3}}$
exhibits symmetries in $(l_{1},l_{2},l_{3})$, the computation time
can be reduced to 1/6. The bispectrum expression (\ref{eq:bispec_simplified})
contains summation over $L_{i}$'s which, in turn, are dependent upon
$l_{i}$'s, as can be seen in Eq. (\ref{eq:ell_conds}). This gives
us the advantage of evaluating $k$ integrals only for valid $(l,L)$
pairs, i.e., Eq. (\ref{eq:the_num_integ}). This was followed by the
$x$ integration. In order to avoid the repetitive calculation of
same $k$ integrals, we saved the integral (\ref{eq:the_num_integ})
information as a function of $x$ on an array. To serve this purpose,
we developed a method for storing and retrieving data, using a binary
search method for integer $n$ tuples. Two theorems provided by us
ensure the certainty of the binary search method. For efficient evaluation,
we first calculated the Wigner symbols by using the open source softwares,
e.g., \texttt{fgsl} and \texttt{wigxjpf}. Then, we stored and retrieve
the required symbols using the \textit{Rasch Algorithm}. We think
that the software \texttt{wigxjpf} used for the evaluation of Wigner
symbols can be coupled with Rasch Algorithm to improve the computation
speed.

The highest absolute value of B-mode bispectrum due to inflation,
as per Fig. \ref{fig:Comparison-of-the-Bispectra}, is $\sim6\times10^{-11}$
which occurs at $(l_{1},l_{2},l_{3})=(2,15,16)$. We find that the
B-mode auto-bispectrum due to matter bounce is about $10^{-12}$ times
smaller as compared to this value and occurs at the same point.  For
numerical integration over $x$ (and $y$) variable(s), we checked that
both the trapezoidal and rectangular integration methods were giving
the same answer.

We also found that the shape functions for the two cases exhibit stark
differences as can be seen in Fig. \ref{fig:Comparison-of-the-Shape}.
We have also calculated the SNR in Figure \ref{fig:Signal-to-Noise}.
For its calculation, we use the same expression as used in Ref. \citep{Tahara2017}.
The SNR for scalar-to-tensor ratio $r=0.01$ and $\l_{\text{max}}=100$
for inflation is $\sim10^{-5}$. For matter bounce, this ratio is
smaller by a factor of $\sim10^{-13}$. This means that it would be
even more difficult to detect signals due to bounce as compared to
inflation. Thus a future detection of tensor bispectrum will  rule out matter bounce model.

\section{Acknowledgments}

We are very much thankful to L. Sriramkumar of the Physics Department,
IIT Madras. We are also indebted to Pankaj Jain of the Physics Department,
IIT Kanpur for reading through the manuscript and suggesting valuable
comments. We are very grateful to Rathul Nath Raveendran, Debika Choudhury
for useful discussions. We wish to thank Indian Institute of Technology
Madras, Chennai, India, for support through the Exploratory Research
Project PHY/17-18/874/RFER/LSRI. RK is thankful to Pradeep Mishra
of Mathematics Department, IIT Madras for vetting the mathematical
appendix. RK also wishes to thank hospitality provided at IIT Kanpur
where computing facilities supported by the Science and Engineering
Research Board (SERB), Government of India were used. Further, we would also like to thank our anonymous referee for encouraging words and pointing one of the most important conclusions of our work that a future detection of tensor bispectrum will be helpful in ruling out matter bounce model. Finally, RK is very thankful to Prof. Roy Maartens for illuminating discussions.

\appendix

\section{\label{sec:Numerical-Strategies}Numerical Strategies}

In this appendix, we give details of the methods that underlie the
numerical evaluation of the bispectrum form in Eq. (\ref{eq:bispec_simplified}).
The evaluation hinges on two important results which we prove next.

\subsection{\label{subsec:Binary-Search}A Binary Search Algorithm for Integer
Tuples}

For numerical evaluation of the integrals in Eq. (\ref{eq:bispec_simplified}),
we stored integrals given in Eqs. (\ref{eq:the_num_integ}) or (\ref{eq:the_num_integ-Lapalce})
on an array, based on whether the bispectrum is separable or inseparable.
The binary search algorithm plays a crucial role in the bispectrum
determination for retrieving these integrals to  perform
integration over $x$ (and $y$) for different values of $l$, $L$
and $n$. It is known that the binary search is one of the fastest
search method when the array is sorted.

Let $X$ denote the finite set of $n$ integers' tuples, i.e.,
\[
X=\left\{ \left.\mathbf{x}^{j}\equiv\left(x_{1},x_{2},\ldots,x_{i}\right)\right|x_{i}\in\mathbb{Z},\ 1\le j\le N\right\} .
\]
Further, let $\mathbf{x}\equiv\left(x_{1},x_{2},\ldots x_{n}\right)$
and $\mathbf{y}\equiv\left(y_{1},y_{2},\ldots,y_{n}\right)$ be two
elements of $X$. We say $\mathbf{x}$ ``comes before'' $\mathbf{y}$
and write $\mathbf{x}\prec\mathbf{y}$ when either of these two cases
are true 
\begin{itemize}
\item[A.] $x_{1}<y_{1}$ 
\item[B.] there exists an $i$ such that if $x_{j}=y_{j}$ for every $1\le j<i$
then $x_{i}<y_{i}$. 
\end{itemize}
Please notice that these are the conditions for dictionary ordering or lexicographic ordering,
the same conditions due to which `car' comes before `cat'. Thus according
to our definition `car' $\prec$ `cat'.

Let $\mathbf{x}^{1},\mathbf{x}^{2},\ldots\mathbf{x}^{k}$ be $k$
number of tuples of $X$, then these tuples are said to be \textit{dictionary
ordered} if there exists a permutation of indices $\left\{ i_{1},i_{2},\ldots i_{n}\right\} $
such that $\mathbf{x}^{i_{1}}\prec\mathbf{x}^{i_{2}}\prec\ldots\prec\mathbf{x}^{i_{k}}$,
where for $1\le p\le k$ we have $1\le i_{p}\le n$. Next is the notion
of \textit{maximal element} which would be defined as $m_{1}=\max\left\{ x_{1}^{i_{k}},x_{2}^{i_{k}},\ldots,x_{n}^{i_{k}}\right\} $
and \textit{minimal element} of $X$ would be defined as $m_{2}=\min\left\{ x_{1}^{i_{1}},x_{2}^{i_{1}},\ldots,x_{n}^{i_{1}}\right\} $.
In simple words, the maximal element is the largest integer of the
tuple that comes last in the lexicographic ordering. Similarly, minimal
is the smallest integer in the tuple that comes first. Notice that
$m_{1}-m_{2}>0$, since if $m_{1}=m_{2}$ then there would only be
one element in $X$.

As an example, consider the tuple set $X=\left\{ \left(1,2,3\right),\left(-1,2,3\right),\left(1,2,5\right)\right\} $.
The corresponding dictionary ordered set would be $\left\{ \left(-1,2,3\right),\left(1,2,3\right),\left(1,2,5\right)\right\} $,
maximal element $m_{1}$ would be $5$ and minimal element $m_{2}$
would be $-1$. The next result describes that there exists a one
to one mapping between $n$ tuple set $X$ and $\mathbb{Z}$.

\label{thm:injective_map} \noindent\textbf{Theorem 1:} Let $X$ be the finite set of $n$ tuples,
then the function $f:X\to\mathbb{Z},\ f\left(\mathbf{a}\right)=\sum_{s=0}^{n-1}M^{s}a_{n-s}$
is one-one, where $M\ge m_{1}+\left|m_{2}\right|+1$. 

\begin{proof}
We will refer to $M$ as the \textit{base of representation}. To show
that the function $f$ is one-one, we demonstrate that 
\[
f\left(\mathbf{a}\right)=f\left(\mathbf{b}\right)\Rightarrow\mathbf{a}=\mathbf{b},
\]
where $\mathbf{a}$, $\mathbf{b}\in X$. Now 
\[
f\left(\mathbf{a}\right)=f\left(\mathbf{b}\right)\Rightarrow\sum_{s=0}^{n-1}M^{s}\left(a_{n-s}-b_{n-s}\right)=0.
\]

For convenience, let us denote $a_{p}-b_{p}=c_{p}$. Notice that $a\mod b$
gives the remainder when $a$ is divided by $b$ \citep{Gallian2016}. We perform the modular
operation $n-1$ number of times with different powers of $M$, \textit{viz.
}$M^{n-1}$, $M^{n-2}$$\ldots,M$ to get the following equations
\[
\begin{bmatrix}M^{n-1} & M^{n-2} & \ldots & M & 1\\
0 & M^{n-2} & \ldots & M & 1\\
\vdots &  &  & \vdots & \vdots\\
0 & 0 & \ldots & M & 1\\
0 & 0 & \ldots & 0 & 1
\end{bmatrix}\begin{bmatrix}c_{1}\\
c_{2}\\
\vdots\\
c_{n-1}\\
c_{n}
\end{bmatrix}=\begin{bmatrix}0\\
0\\
\vdots\\
0\\
0
\end{bmatrix}.
\]

We notice that the matrix is upper triangular and the value of the
determinant is $M^{n\left(n-1\right)/2}\ne0$ if $M\ne0$. Thus by
the Cramer's rule (see \citep{hoffman1971}), the given homogeneous
system has only a trivial solution $\mathbf{c}=0$, thereby implying
$\mathbf{a}=\mathbf{b}$. Therefore the given function is one-one. 
\end{proof}

The result, we just proved shows that the function $f$ is one-one.
If we restrict the range to $f\left(X\right)=D$, then the function
$f:X\to D$ becomes bijective. We will call the function $f$ an ``indexing
scheme.'' The next result shows that if a given tuple array is in
the lexicographic order and the above indexing scheme is used, resulting
array of integers is automatically sorted, i.e., 
\begin{equation}
\mathbf{x}^{i_{1}}\prec\mathbf{x}^{i_{2}}\prec\ldots\prec\mathbf{x}^{i_{k}}\Rightarrow f\left(\mathbf{x}^{i_{1}}\right)<f\left(\mathbf{x}^{i_{2}}\right)<\ldots<f\left(\mathbf{x}^{i_{k}}\right).
\end{equation}

This means that one need not use any sorting methods and hence one
can directly use the binary search method. The proof follows next. 

\label{thm:lexicographic} \noindent\textbf{Theorem 2:} Let $\mathbf{a}$ and $\mathbf{b}$ be
two $n$ tuples of $X$ then $\mathbf{a}\succ\mathbf{b}\Rightarrow f\left(\mathbf{a}\right)>f\left(\mathbf{b}\right)$. 

\begin{proof}
As per the definition of `$\succ$', we need to consider two cases
for $\mathbf{a}\succ\mathbf{b}$: 
\begin{itemize}
\item[1.] $b_{1}<a_{1}$ 
\item[2.] there exists an $i$ such that if $a_{j}=b_{j}$ for every $1\le j<i$
then $b_{i}<a_{i}$. 
\end{itemize}
Let us analyze Case 1 first. We calculate the quantity $f\left(\mathbf{a}\right)-f\left(\mathbf{b}\right)$
which is equal to 
\begin{equation}
\sum_{s=0}^{n-1}M^{s}\left(a_{n-s}-b_{n-s}\right)=M^{n-1}\left(a_{1}-b_{1}\right)+\sum_{s=0}^{n-2}M^{s}\left(a_{n-s}-b_{n-s}\right).
\end{equation}

But $a_{i}$ and $b_{i}$ are integers therefore $a_{1}>b_{1}\Rightarrow a_{1}-b_{1}\ge1$
and since $m_{1}$ and $m_{2}$ are the maximal and minimal elements
of $X$ therefore $m_{2}-m_{1}\le a_{i}-b_{i}$ for every $1\le i\le n$.
Thus we must have 
\[
M^{n-1}+\left(m_{2}-m_{1}\right)\left(\frac{M^{n-1}-1}{M-1}\right)\le f\left(\mathbf{a}\right)-f\left(\mathbf{b}\right).
\]
The second term is obtained by summing over a geometric series. For
further analysis, notice that $M\ge m_{1}+\left|m_{2}\right|+1\ge m_{1}-m_{2}+1$
so that 
\begin{equation}
	\frac{m_{2}-m_{1}}{M-1}\ge-1\Rightarrow\left(\frac{m_{2}-m_{1}}{M-1}\right)\left(M^{n-1}-1\right)\ge-\left(M^{n-1}-1\right).
	\end{equation}
So we have 
\begin{equation}
M^{n-1}-\left(M^{n-1}-1\right)  \le M^{n-1}+\left(m_{2}-m_{1}\right) \left(\frac{M^{n-1}-1}{M-1}\right) \le f\left(\mathbf{a}\right)-f\left(\mathbf{b}\right).
\end{equation}

From this, we get the required inequality $f\left(\mathbf{a}\right)>f\left(\mathbf{b}\right)$.
Now let us consider the second Case. In this case, we can write $f\left(\mathbf{a}\right)-f\left(\mathbf{b}\right)$
as 
\begin{equation}
\sum_{s=0}^{n-1}M^{s}\left(a_{n-s}-b_{n-s}\right)  =M^{n-i}\left(a_{i}-b_{i}\right)
  +\sum_{s=n-i+1}^{n-1}M^{s}\left(a_{n-s}-b_{n-s}\right).
\end{equation}
This follows since $a_{j}=b_{j}$ when $1\le j<i$. Using the similar
reasoning as was done for Case 1 we get 
\begin{equation}
M^{n-i}-\left(M^{n-i}-1\right) \le M^{n-i}+\left(m_{2}-m_{1}\right)\frac{M^{n-i}-1}{M-1}
 \le f\left(\mathbf{a}\right)-f\left(\mathbf{b}\right).
\end{equation}
Thus again the inequality $f\left(\mathbf{a}\right)>f\left(\mathbf{b}\right)$
is satisfied. 
\end{proof}

\section{\label{sec:Simplification-of-Bispectrum-Matter}Simplification of
Bispectrum due to Matter Bounce}

The function $\mathcal{G}\left(k_{1},k_{2},k_{3}\right)$ appearing
in Eq. (\ref{eq:f_func}) for matter bounce can be written as \citep{Chowdhury2015}
\begin{equation}
\mathcal{G}\left(k_{1},k_{2},k_{3}\right)=\left(2\pi\right)^{-17/2}\,8M_{\text{Pl}}^{2}\left(\mathbb{G}\left(k_{1},k_{2},k_{3}\right)+\text{c.c.}\right),\label{eq:g_func_bounce}
\end{equation}
where $M_{\text{Pl}}=10^{56}\,{\rm Mpc}^{-1}$ represents the Planck
mass expressed in the Mpc units, `c.c.' represents the complex conjugation
of the quantity coming before it. In Ref. \citep{Chowdhury2015},
in order to evaluate the tensor bispectra, authors divided the time
regime into three domains and showed that, only in the third domain,
it contributes the most. In that domain, $\mathbb{G}(k_{1},k_{2},k_{3})$
takes the following form 

\begin{eqnarray}
\mathbb{G}\left(k_{1},k_{2},k_{3}\right)  &=& \left[A_{k_{1}}^{*}A_{k_{2}}^{*}A_{k_{3}}^{*}J_{0}\left(\beta\right)+\text{Perm}\left(A_{k_{1}}^{*}A_{k_{2}}^{*}B_{k_{3}}^{*}\right)J_{1}\left(\beta\right)
+\text{Perm}\left(A_{k_{1}}^{*}B_{k_{2}}^{*}B_{k_{3}}^{*}\right)J_{2}\left(\beta\right)+\right.\nonumber\\&&\left.\qquad\qquad\qquad\qquad\qquad\qquad B_{k_{1}}^{*}B_{k_{2}}^{*}B_{k_{3}}^{*}J_{3}\left(\beta\right)\right]
\times\left(-\frac{ia_{0}}{4k_{0}}\right)h_{k_{1}}h_{k_{2}}h_{k_{3}}.
\end{eqnarray}

In this equation, $a_{0}=10^{-30}$ is the minimum value of the scale
factor that the Universe takes at the bounce, $k_{0}^{-1}=10^{-20}{\rm Mpc}$
is the energy scale that determines the duration of bounce. Other
quantities are defined in the following manner:

\begin{equation}
h_{k}=A_{k}+B_{k}f\left(\eta\right),\ f\left(x\right)=\frac{x}{1+x^{2}}+\tan^{-1}\left(x\right),\label{eq:func_fx}
\end{equation}
with
\begin{eqnarray}
B_{k} &=&\frac{\left(1+\alpha^{2}\right)^{2}}{2a_{0}\alpha^{2}M_{\text{pl}}\sqrt{k}}\left[\frac{3ik_{0}}{\alpha^{2}k}+\frac{3}{\alpha}-\frac{ik}{k_{0}}\right]\exp\left(\frac{i\alpha k}{k_{0}}\right),\quad \\
A_{k} &=&\frac{1}{a_{0}\alpha^{2}M_{\text{pl}}\sqrt{k}}\left(1+\frac{ik_{0}}{\alpha k}\right)\exp\left(\frac{i\alpha k}{k_{0}}\right)+f\left(\alpha\right)B_{k},
\end{eqnarray}

In Eq. (\ref{eq:func_fx}), $\eta$ is the comoving time at which
the calculation is performed. The constant $\alpha=10^{5}$ is chosen
so that the relevant comoving length scales $k\ll k_{0}/\alpha$ and
that the tensor power spectrum becomes scale invariant. The constants
$J_{i}\left(\beta\right)$ are defined through an integral 
\[
J_{n}\left(x\right)=\int_{0}^{x}\left(1+y^{2}\right)^{2}f^{n}\left(y\right),
\]
function $f\left(y\right)$ being given in Eq. (\ref{eq:func_fx}).
The symbol $\text{Perm}\left(A_{k_{1}}B_{k_{2}}C_{k_{3}}\right)$
denotes all possible permutations of the terms inside the bracket
with the given subscripts. For example 
\begin{equation}
	\text{Perm}\left(A_{k_{1}}A_{k_{2}}B_{k_{3}}\right)=A_{k_{1}}A_{k_{3}}B_{k_{2}}+A_{k_{1}}A_{k_{3}}B_{k_{1}}
	+A_{k_{1}}A_{k_{2}}B_{k_{3}}.\nonumber
\end{equation}
Simplification of the term inside parenthesis of Eq. (\ref{eq:g_func_bounce})
becomes the following:

\begin{align}
\mathbb{G}+\mathbb{G}^{*} & =\left[\left(cJ_{0}-J_{1}\right)\text{Perm}\left(\Lambda_{\lambda_{1}}\phi_{\lambda_{2}}\Lambda_{\lambda_{3}}\right)+\left(c^{2}J_{0}-J_{2}\right)\text{Perm}\left(\theta_{\lambda_{1}}\Lambda_{\lambda_{2}}\phi_{\lambda_{3}}\right)+\left(c^{3}J_{1}-cJ_{3}\right)\text{Perm}\left(\theta_{\lambda_{1}}\Gamma_{\lambda_{2}}\phi_{\lambda_{3}}\right)\right.\nonumber \\
 & +\left(c^{2}J_{1}-cJ_{2}\right)\text{Perm}\left(\Lambda_{\lambda_{1}}\Gamma_{\lambda_{2}}\phi_{\lambda_{3}}\right)+\left(c^{2}J_{1}-cJ_{2}+c^{3}J_{0}-J_{3}\right)\text{Perm}\left(\theta_{\lambda_{1}}\theta_{\lambda_{2}}\phi_{\lambda_{3}}\right)\nonumber \\
 & \left.+\left(3c^{2}J_{1}-3cJ_{2}-c^{3}J_{0}+J_{3}\right)\phi_{1}\phi_{2}\phi_{3}+\left(c^{3}J_{2}-c^{2}J_{3}\right)\text{Perm}\left(\Gamma_{\lambda_{1}}\Gamma_{\lambda_{2}}\phi_{\lambda_{3}}\right)\right],\label{eq:G+ccG}
\end{align}

where the functions are defined below. 
\begin{eqnarray*}
\Lambda &=&  \left|A\right|^{2} = \frac{1}{(2k_{0}M_{\text{pl}}^{4}a_{0}^{5})^{1/3} k\alpha^{4}}\left[\left(1+\frac{3f\left(\alpha\right)\left(1+\alpha^{2}\right)^{2}}{2\alpha}\right)^{2}  +\left\{ \frac{k_{0}}{\alpha k}+\frac{f\left(\alpha\right)\left(1+\alpha^{2}\right)^{2}}{2}\left(\frac{3k_{0}}{\alpha^{2}k}-\frac{k}{k_{0}}\right)\right\} ^{2}\right],\label{eq:lambda} \\
\Gamma &=& \left|B\right|^{2}=\frac{\left(\alpha^{-1}+\alpha\right)^{4}}{(M_{\text{pl}}^4k_{0}2^7a_{0}^{5})^{1/3}k}\left[\frac{9}{\alpha^{2}}+\left(\frac{3k_{0}}{\alpha^{2}k}-\frac{k}{k_{0}}\right)^{2}\right], \\
\phi &=& {\rm Re}\left(AB^{*}\right)=\frac{\left(1+\alpha^{2}\right)^{2}}{(2^{4}k_{0}^4M_{\text{pl}}^4a_{0}^5)^{1/3}\alpha^{4}},\label{eq:phi}
\\
\theta &=& {\rm Im}\left(AB^{*}\right) = \frac{\left(1+\alpha^{2}\right)^{2}}{(M_{\text{pl}}^4k_{0}a_{0}^52^4)^{1/3}k\alpha^{4}}\left[\frac{2}{\alpha}+\frac{3k_{0}^{2}}{\alpha^{3}k^{2}}+\frac{f\left(\alpha\right)\left(1+\alpha^{2}\right)^{2}}{2}\left\{ \left(\frac{3k_{0}}{\alpha^{2}k}-\frac{k}{k_{0}}\right)^{2}+\frac{9}{\alpha^{2}}\right\} \right].\label{eq:theta}
\end{eqnarray*}

These functions can be further simplified, owing to the values of
the various parameters of the theory and the following final form
is obtained 
\begin{eqnarray}
\Lambda\approx\frac{9\pi^{2}C_{1}}{2^{1/3}k^{3}},\quad\Gamma\approx\frac{9C_{1}}{2^{7/3}k^{3}},\quad\theta\approx\frac{9C_{1}}{2^{7/3}k^{3}},\quad \phi\approx\frac{C_{2}}{2^{4/3}},\label{eq:C_1}
\end{eqnarray}

The constants $C_{1}, C_2$ appearing in Eqs. (\ref{eq:C_1}) are: 
\begin{eqnarray}
C_{1} &=& \left(\frac{k_{0}^{5}}{M_{\text{pl}}^{4}a_{0}^{5}}\right)^{1/3}=\left(10^{26}{\rm Mpc}^{-1}\right)^{1/3},\\
C_{2} &=& \left(\frac{1}{M_{\text{pl}}^{4}a_{0}^{5}k_{0}^{4}}\right)^{1/3}=\left(10^{-154}{\rm Mpc}^{-8}\right)^{1/3}.
\end{eqnarray}
Since $C_{2}$ appears with $\phi$ term, this means that the $\phi\phi$
and other higher order terms will be suppressed, due to which,
the function $\mathbb{G}$ of Eq. \ref{eq:G+ccG} can finally be written
as 
\begin{equation}
	\mathbb{G}=C\left[\frac{1}{\left(k_{1}k_{2}\right)^{3}}+\frac{1}{\left(k_{2}k_{3}\right)^{3}}+\frac{1}{\left(k_{3}k_{1}\right)^{3}}\right],
\end{equation}
the constant $C$ being 
\begin{eqnarray}
C &=&\frac{72\times10^{-34}}{\pi^{6}5\sqrt{7}}[8\pi^{2}\left\{ 2\pi^{2}\left(cJ_{0}-J_{1}\right)+c^{2}J_{0}+c^{2}J_{1} -J_{2}\left(c+1\right)\right\}\nonumber\\
&&  +\left(2c^{3}+c^{2}\right)J_{1}+c^{3}J_{0}+J_{2}\left(c^{3}-c\right)-J_{3}\left(1+2c+c^{3}\right)],\label{eq:constant}
\end{eqnarray}
and $c=f\left(\beta\right)$. 

\providecommand{\href}[2]{#2}\begingroup\raggedright\endgroup

\end{document}